\documentclass[12pt,a4paper]{article}
\usepackage{amsmath,amssymb,amsthm}
\usepackage{array,graphicx}
\usepackage{fullpage}

\numberwithin{equation}{section}

\newcommand{\mysection}[1]{%
    \addtocounter{MySection}{1}\setcounter{equation}{0}%
    {\bf\arabic{MySection}. #1.}%
}
\begin{document}
\graphicspath{{C:/Sergey/Papers/2008/Pictures/}}
\newcounter{MySection}
\begin{center}

\bf
EXACT SOLUTION DESCRIBING A SHALLOW\\[1mm] WATER FLOW IN AN EXTENDING STRIPE

\bigskip

Sergey V. Golovin

\bigskip
\rm 
Lavrentyev Institute of Hydrodynamics of SB RAS,\\ Novosibirsk State University\\
Novosibirsk, 630090, Russia\\
e-mail: sergey@hydro.nsc.ru
\end{center}

Partially invariant solution to $(2+1)$D shallow water equation is constructed and investigated. The solution describes an extension of a stripe, bounded by linear source and drain of fluid. Realizations of smooth flow and of hydraulic jump are possible. Particle trajectories and sonic characteristics on the obtained solution are calculated.

\mysection{Construction of the solution} Equations describing a thin layer of water flows over a flat bottom are observed.
\begin{equation}\label{shall2D}
\begin{array}{l}
u_t+uu_x+vu_y+h_x=0,\\[1mm]
v_t+uv_x+vv_y+h_y=0,\\[1mm]
h_t+(uh)_x+(vh)_y=0.
\end{array}
\end{equation}
Here $(u,v)$ is a velocity vector of a particle, $h$ is a depth of fluid. The gravity acceleration is scaled to $g=1$. Equations (\ref{shall2D}) admit 9-dimensional Lie algebra of infinitesimal transformations $L_9$ \cite{LVO1}, generated by operators (notations of \cite{Pavl} are adopted)
\[\begin{array}{l}
X_1=\partial_x,\quad X_2=\partial_y,\quad X_4=t \partial_x +\partial_u,\quad X_5=t\partial_y +\partial_v,\\[1mm]
X_9=x \partial_y-y\partial_x+u\partial_v-v\partial_u,\quad X_{10}=\partial_t, \\[1mm]
X_{11}=x\partial_x+y\partial_y+u\partial_u+v\partial_v+2h\partial_h,\\[1mm]
X_{12}=t^2\partial_t+tx\partial_x+ty\partial_y+(x-tu)\partial_u+(y-tv)\partial_v-2th\partial_h.\\[1mm]
X_{13}=2t\partial_t+x\partial_x+y\partial_y-u\partial_u-v\partial_v-2h\partial_h.
\end{array}\]
We observe a partially invariant solution generated by Lie subalgebra
\[N=\{X_1,\; X_4,\; X_{10}+X_{12}\}\subset L_9.\]
Invariants of $N$ are the following functions
\[ \lambda=\frac{y}{\sqrt{t^2+1}},\quad V=v\sqrt{t^2+1}-t\lambda,\quad
   H=h(t^2+1).\]
According to the general algorithm \cite{LVO} the representation of $N$-partially invariant solution can be written as
\[ V=V(\lambda),\quad H=H(\lambda),\quad u=u(t,x,y). \]
There are two invariant functions $V$ and $H$, which are set to depend on the only invariant variable $\lambda$, and one non-invariant function $u$, which depend on all independent variables. We substitute an expressions for invariants $V$ and $H$ in this representation, and then solve it with respect to unknown functions $v$ and $h$. In what follows it is convenient to represent function $u$ as a function of $(t,x,\lambda)$ instead of  $(t,x,y)$. These give the following formulae:
\begin{equation}\label{NHrepr}
u=u(t,x,\lambda),\quad v=\frac{V(\lambda)+t\lambda}{\sqrt{t^2+1}},
   \quad h=\frac{H(\lambda)}{t^2+1},\;\;\;\lambda=\frac{y}{\sqrt{t^2+1}}.
\end{equation}
Substitution of the representation \eqref{NHrepr} into \eqref{shall2D} gives
\begin{equation}\label{shall2D_1}
\begin{array}{l}
u_t+uu_x+V(t^2+1)^{-1}u_\lambda=0,\;\\[2mm]
VV'+H'=-\lambda,\\[2mm]
VH'+H\bigl(V'+(t^2+1)u_x-t\bigr)=0.
\end{array}
\end{equation}
Here the lower indexes denote partial derivatives with respect to the corresponding variable; prime is the derivative with respect to  $\lambda$. The first and the third equations of system \eqref{shall2D_1} form an overdetermined system  $\Pi$ for non-invariant function $u$.

At first, let us start from the subsystem $\Pi$. Separation of variables in the third equation of
\eqref{shall2D_1} allows introduction of the new invariant function
\[K(\lambda)=-VH'/H-V'.\]
The non-invariant function $u$ should satisfy the subsystem:
\begin{equation}\label{shall2D_2}
u_t+uu_x+\frac{V}{t^2+1}u_\lambda=0,\quad (t^2+1)u_x-t=K.
\end{equation}
Integration of the second equation of \eqref{shall2D_2} gives
\begin{equation}\label{shall2DNonInv}
u=x\frac{K(\lambda)+t}{t^2+1}+U(t,\lambda).
\end{equation}
Substitution of this representation into the first equation of \eqref{shall2D_2} produces a compatibility condition in form of an equation for $K$:
\[ VK'+K^2+1=0 \]
and a relation for function $U(t,\lambda)$:
\begin{equation}\label{shall2D_3}
(t^2+1)U_t+VU_\lambda+(K+t)U=0.
\end{equation}
Thus, the  factor-system \eqref{shall2D_1} splits into the invariant subsystem
\begin{equation}\label{shall2D_4}
\begin{array}{l}
VV'+H'=-\lambda,\\
VK'+K^2+1=0,\\
VH'+HV'=-KH.
\end{array}
\end{equation}
and equation \eqref{shall2D_3} for function $U$. These equations admit a discrete symmetry
\begin{equation}\label{discrsymm}
\lambda\to-\lambda,\;\;\;V\to-V,
\end{equation}
which follows from the admissible by equations (\ref{shall2D}) transformation $y\to-y$, $v\to-v$. Integration of the obtained nonlinear system of differential equations can be performed in terms of the new independent variable $\mu$:
\begin{equation}\label{shall2D_5}
\frac{d\lambda}{d\mu}=V(\lambda),\quad \mu=\int\frac{d\lambda}{V(\lambda)}.
\end{equation}
With the new variable, the second equations of \eqref{shall2D_4} accurate to insufficient constant gives
\begin{equation}\label{shall2D_6}
K=-\tan \mu.
\end{equation}
Taking \eqref{shall2D_6} into account, one can integrate equation \eqref{shall2D_3} as
\[ U=\frac{f(\mu-\arctan t)}{\cos\mu\sqrt{t^2+1}} \]
($f$ is an arbitrary function). Besides, system \eqref{shall2D_4} has a first integral, which follows from its third equations:
\begin{equation}
HV\cos\mu=m,\;\;\;m=\mathrm{const}.
\end{equation}
According to the discrete symmetry (\ref{discrsymm}) it is enough to observe only positive values of $V$. Invariant depth $H$ in non-negative by its physical definition, therefore, $m\cos\mu>0$. As $\cos \mu\ne0$, the admissible set of values of functions and parameters can always be restricted to the following:
\begin{equation}\label{values}
\lambda\in\mathbb{R}, \;\;\; -\frac{\pi}{2}<\mu<\frac{\pi}{2},\;\;\;V>0,\;\;\;\lambda'(\mu)>0,\;\;\;m>0.
\end{equation}
Finally, equations \eqref{shall2D_4} possess a Bernoulli integral:
\begin{equation}\label{Bern}
V^2+\lambda^2+2H=b^2
\end{equation}
with an arbitrary constant $b$. This integral should be observed as an implicit (not resolved with respect to the derivative) ordinary differential equation for the dependence $\lambda(\mu)$:
\begin{equation}\label{ImplEqn}
\bigl(\lambda'\bigr)^2+\lambda^2+\frac{2m}{\lambda'\cos\mu}=b^2.
\end{equation}
Thus, the solution is given by expressions (\ref{NHrepr}), (\ref{shall2DNonInv}), in which functions $V$, $H$, $K$, and $U$ can be determined from (\ref{shall2D_5})--(\ref{Bern}) after the first-order ODE (\ref{ImplEqn}) integration. This solution contain an arbitrary function $f$ in the expression for the velocity component $u$. Equation \eqref{ImplEqn} plays the key role in the further investigations.

\begin{figure}
\begin{minipage}[b]{0.45\textwidth}
\includegraphics[width=\textwidth]{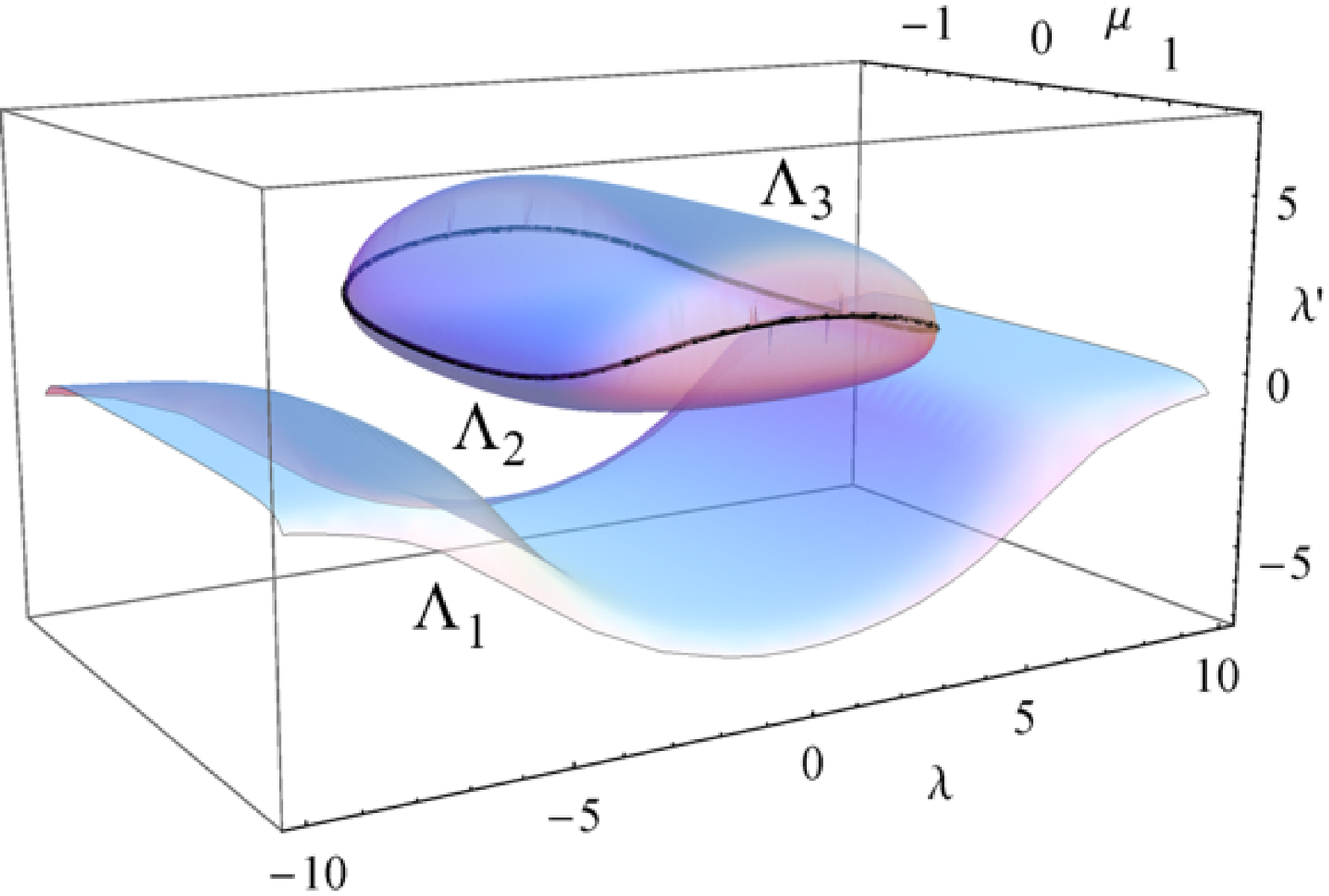}
\caption{Representation of the implicit equation (\ref{ImplEqn}) as a smooth manifold $\Lambda=\Lambda_1\cup\Lambda_2\cup\Lambda_3$ in an extended space $\mathbb{R}^3(\mu,\lambda,\lambda')$.}\label{f1}
\end{minipage}
\hfill
\begin{minipage}[b]{0.45\textwidth}
\includegraphics[width=0.9\textwidth]{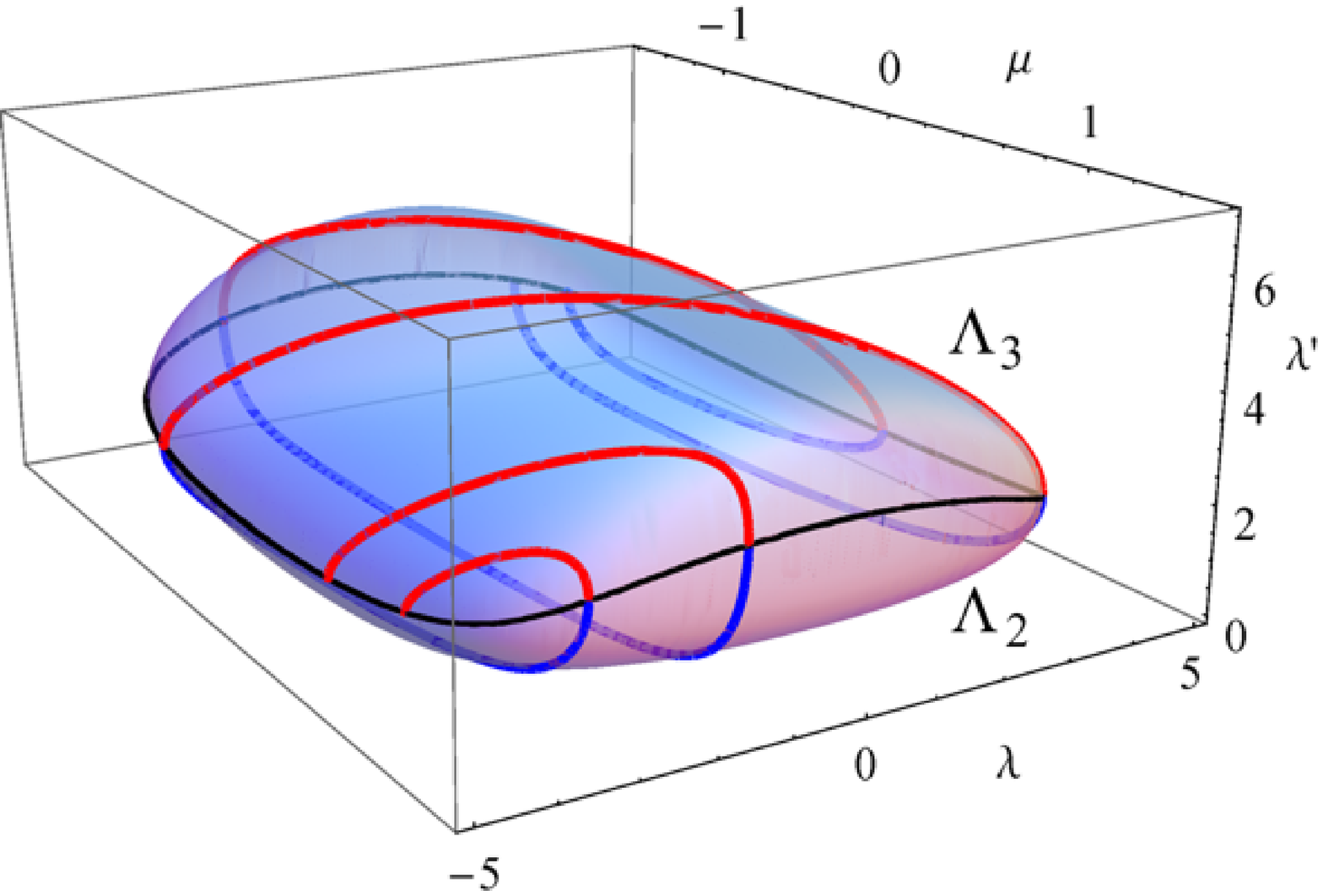}
\caption{Integral curves of equation (\ref{ImplEqn}) on the surface $\Lambda_2\cup\Lambda_3$. The criminant curve, separating $\Lambda_2$ and $\Lambda_3$, divide integral curves into sub- and supercritical parts. }\label{f2}
\end{minipage}
\end{figure}

\mysection{Properties of solutions of the key equation}
For the analysis of equation \eqref{ImplEqn} it is convenient to observe a surface $\Lambda$, defined by the relation (\ref{ImplEqn}) in an extended space $\mathbb{R}^3(\mu,\lambda,\lambda')$ (see fig. \ref{f1}). Depending on the sign of the discriminant
\begin{equation}\label{Ddef}
{\cal D}=\frac{m^2}{\cos^2\mu}-\left(\frac{b^2-\lambda^2}{3}\right)^3
\end{equation}
the cubic algebraic equation (\ref{ImplEqn}) for the derivative $\lambda'$ possesses 1, 2 or 3 real roots. The surface $\Lambda$ is correspondingly consists of three components $\Lambda=\Lambda_1\cup\Lambda_2\cup\Lambda_3$ responsible for each of three roots. The negative root $\lambda_1'<0$ exists for any value of parameters; the corresponding unlimited component of the manifold $\Lambda$ is $\Lambda_1$ (see fig. \ref{f1}). The remaining two components $\Lambda_2$ and $\Lambda_3$ corresponding to positive roots $0<\lambda_2'<\lambda_3'$ are finite. They close up along the so-called criminant curve specified by an additional relation ${\cal D}=0$. Physical meaning have only positive roots, since the negative root $\lambda_1'=V<0$ does not satisfy conditions \eqref{values} (the negative root corresponds to the negative water depth $h$). Surface $\lambda$ is woven from integral curves of equation \eqref{ImplEqn} (see fig. \ref{f2}). In accordance to the general theory of equations, not resolved with respect to the highest derivative \cite{Arnold}, through each nonsingular point of the smooth surface $\Lambda\subset\mathbb{R}^3(\mu,\lambda,\lambda')$ passes only one integral curve of the equation. The singular point could appear only at the criminant curve ($=$ intersection of the cylinder ${\cal D}=0$ with the surface $\Lambda_2\cup\Lambda_3$). Solutions of the equation \eqref{ImplEqn} in form $\lambda=\lambda(\mu)$ are obtained after the projection of three-dimensional integral curves into $(\mu,\lambda)$-plane.

\begin{figure}
\begin{center}
\includegraphics[width=0.5\textwidth]{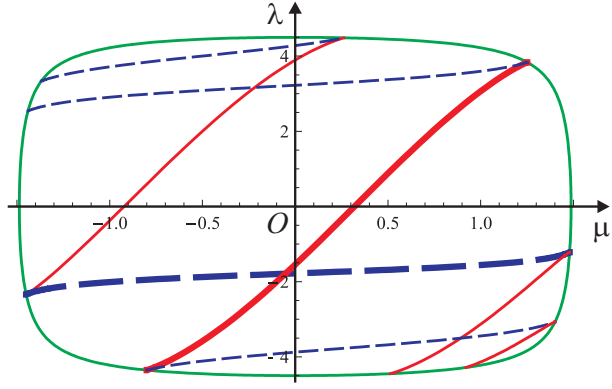}
\caption{Integral curves of equations (\ref{ImplEqn}) for $m=b=2$. The domain of integral curves is limited by the discriminant curve. The solid (dashed) curves correspond to the larger (smaller) positive root $\lambda'$ of the cubic equation (\ref{ImplEqn}). Bold integral curves satisfy boundary conditions $\mu_\ast=-1.450$, $\lambda_\ast=-2.341$, $\mu^\ast=1.480$, $\lambda^\ast=-1.217$ (dashed line), and $\mu_\ast=-0.8$, $\lambda_\ast=-4.352$, $\mu^\ast=1.251$, $\lambda^\ast=3.833$ (solid line). }\label{f3}
\end{center}
\end{figure}

Thus, through each point of an area $\{(\mu,\lambda)\;|\;{\cal D} (\mu,\lambda)<0\}$ pass two physically approved integral curves: one from the projection of $\Lambda_2$, and another from the projection of $\Lambda_3$. Inside of the area the integral curves have no singularities, therefore they can be found with any desired precision by the numerical integration of the equation \eqref{ImplEqn} after its explicit resolution for the derivative $\lambda'$. The picture of the resulting integral curves is given in figure \ref{f3}. In what follows the fluid flow is referred to as the subcritical one if the corresponding integral curve of equation \eqref{ImplEqn} lies entirely in $\Lambda_2$ surface; and as supercritical one if the integral curve lies in $\Lambda_3$. Reasoning for this definition will be given below.

\begin{figure}
\includegraphics[width=0.45\textwidth]{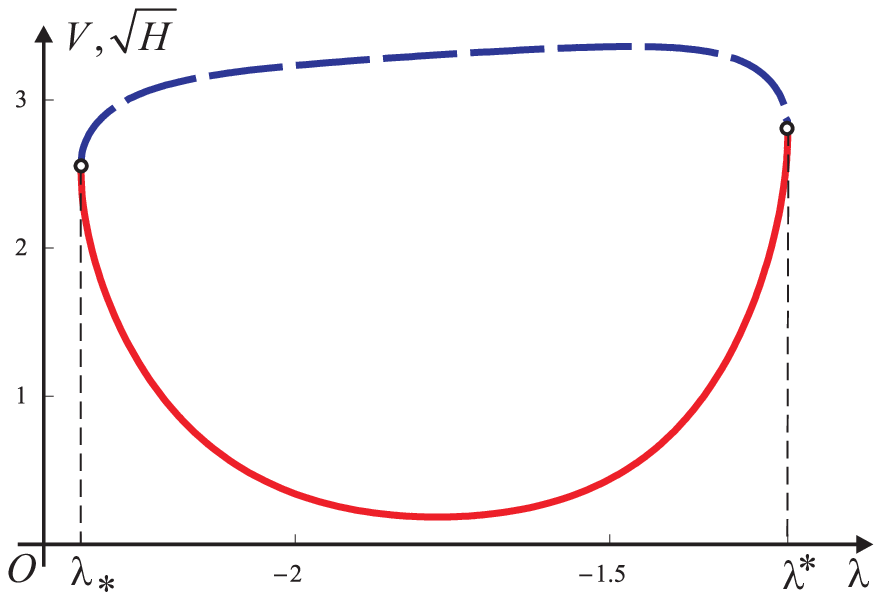}\hfill
\includegraphics[width=0.45\textwidth]{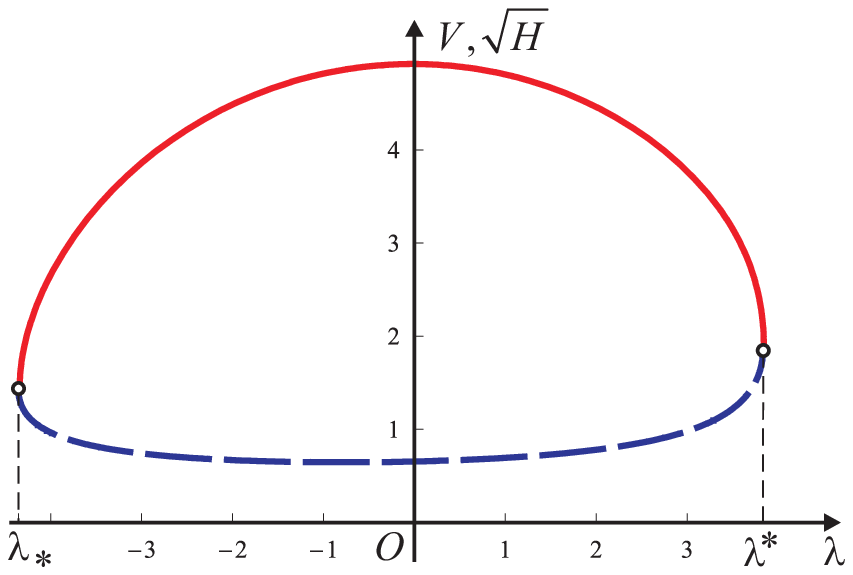}
\caption{Dependencies $V(\lambda)$ (solid line) and $\sqrt{H(\lambda)}$ (dashed line) for subcritical (left-hand figure) and supercritical (right-hand figure) flows. Calculations are made for the integral curves shown bold in figure \ref{f3}.}\label{f45}
\end{figure}

Domain of each integral curve is the finite interval $\mu_\ast\le\mu\le\mu^\ast$ (values $\mu_\ast$, and $\mu^\ast$ are unique for each integral curve). At the boundaries of the domain (at the points of the discriminant curve ${\cal D}=0$) subcritical and supercritical integral curves close up. At that, functions $V(\lambda)$, and $H(\lambda)$ are finite:
\begin{equation}\label{Vlim}
V(\lambda_\ast)=\sqrt{H(\lambda_\ast)}=\left(\frac{m}{\cos\mu_\ast}\right)^{1/3},\;\;\;
V(\lambda^\ast)=\sqrt{H(\lambda^\ast)}=\left(\frac{m}{\cos\mu^\ast}\right)^{1/3}.
\end{equation}
However, their derivatives go into infinity. Indeed, by the Chain Rule $V'(\lambda)=\lambda''(\mu)/\lambda'(\mu)$. Equation \eqref{ImplEqn} gives
\[\lambda''(\mu)=\frac{2\lambda'^2+2m\sin\mu(\cos\mu)^{-2}}{3\lambda'^2+\lambda^2-b^2}.\]
Positive roots of the cubic equation \eqref{ImplEqn} coincide over the discriminant curve: $\lambda_2'=\lambda_3'$. Hence, according to Vi\`{e}te's theorem, $3\lambda_2'^2=b^2-\lambda^2$, i.e. the denominator of the fraction is zero. Therefore, $\lambda''(\mu_\ast)=\lambda''(\mu^\ast)=\infty$. Graphics of dependencies $V(\lambda)$, and $\sqrt{H(\lambda)}$ for sub- and supercritical flows are given in figure \ref{f45}.

\begin{figure}
\includegraphics[width=0.49\textwidth]{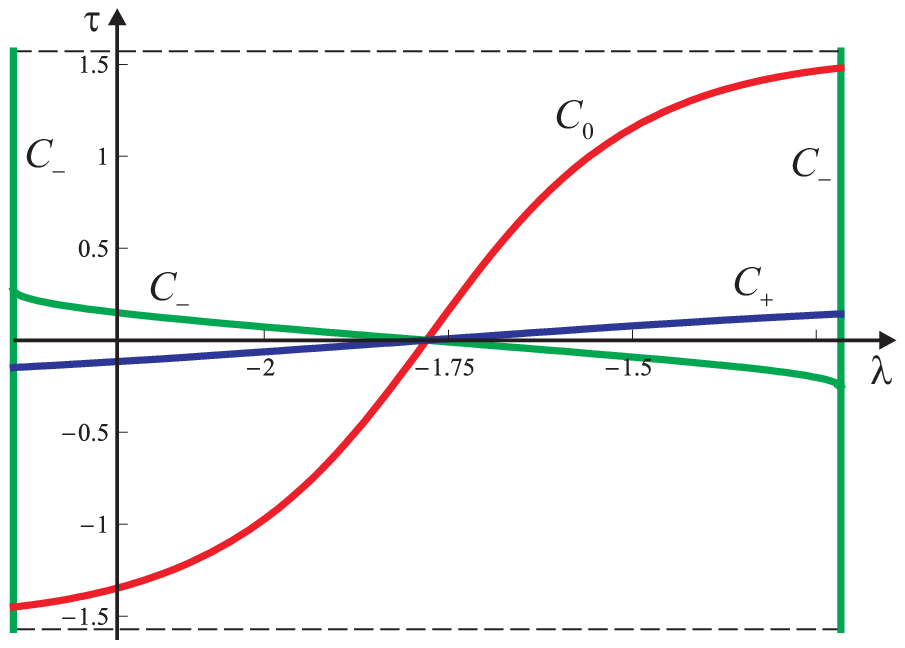}\hfill
\includegraphics[width=0.45\textwidth]{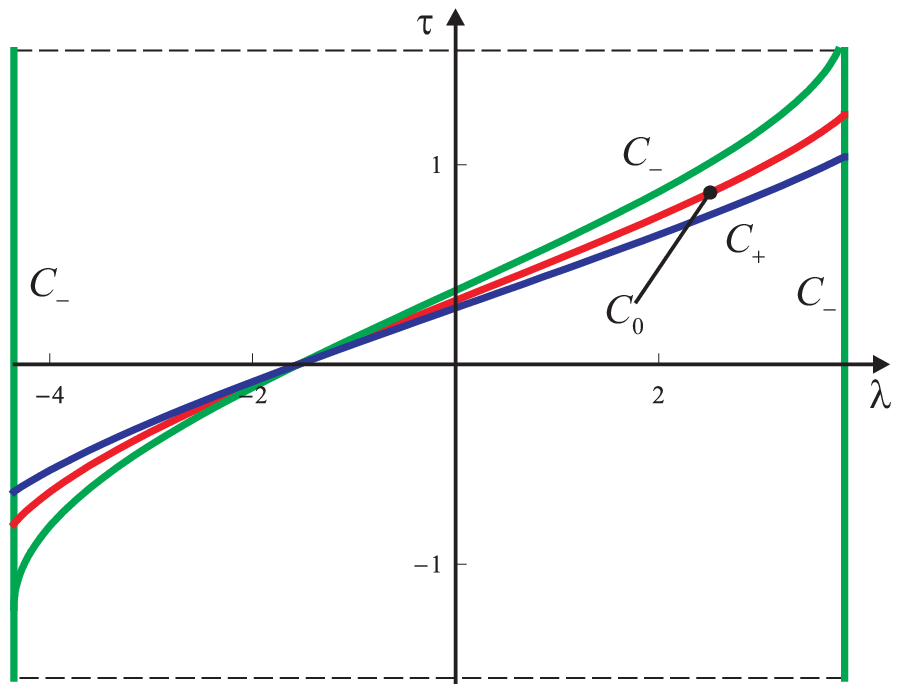}
\caption{Characteristic curves of the subcritical (left-hand figure) and supercritical (right-hand figure) flows, for integral curves of equation \eqref{ImplEqn} shown bold in figure \ref{f3}.}\label{f67}
\end{figure}

\mysection{Characteristics of the shallow water equations on the obtained solution}
Let the characteristic surface of equations \eqref{shall2D} be sought in the implicit form $\chi(t,x,y)=\mathrm{ const}$. There are three families of characteristics: contact ones
\[C_0:\chi_t+u\chi_x+v\chi_y=0\]
and two sonic ones
\[C_\pm:\chi_t+u\chi_x+v\chi_y=\pm\sqrt{h}\sqrt{\chi_x^2+\chi_y^2}.\]
Let us investigate the set of characteristics, specified by the equation $\lambda=\lambda(t)$. Substitution of representations \eqref{NHrepr}, \eqref{shall2DNonInv} gives equations for function $\lambda(t)$ along each of characteristics families in the form
\begin{equation}\label{charact}
C_0:\;\frac{d\lambda}{dt}=\frac{V(\lambda)}{t^2+1},\;\;\;C_\pm:\;
\frac{d\lambda}{dt}=\frac{V(\lambda)\pm\sqrt{H(\lambda)}}{t^2+1}.
\end{equation}
In terms of new variables $\mu$ and $\tau=\arctan t$ equations \eqref{charact} can be integrated as
\begin{equation}\label{charact1}
C_0:\;\tau-\tau_0=\mu(\lambda),\;\;\;C_\pm:\;
\tau-\tau_0=\int\limits_{y_0}^\lambda\frac{d\lambda}{V(\lambda)\pm\sqrt{H(\lambda)}}.
\end{equation}
Note, that for any chosen integral curve of equation \eqref{ImplEqn} surfaces $\lambda=\lambda_\ast$ and $\lambda=\lambda^\ast$ in the space $\mathbb{R}^3(t,x,y)$ are sonic characteristics belonging to $C_-$ family. They are enveloping surfaces for the rest of characteristics of $C_-$ family, given by equation \eqref{charact1}.  The typical characteristic curves \eqref{charact1} in $(\lambda,\tau)$-plane are shown in figure \ref{f67}. All the remaining ones are obtained by shifting curves in figure \ref{f67} along $O\tau$-axis. In subcritical flow the disturbances velocity is higher then the fluid velocity, therefore, the disturbances are transmitted upstream reaching the left boundary $\lambda=\lambda_\ast$ of the domain of the solution. In the supercritical flow the situation is opposite: All the disturbances descend down the flow such that at the left boundary $\lambda=\lambda_\ast$ all characteristics are outgoing.

\mysection{Hydraulic jump}
Let us show that the hydraulic jump may switch the supercritical flow to the subcritical one. Let the position of the hydraulic jump be fixed as $\lambda=\lambda_0$ or, in physical variables, as $y(t)=\lambda_0\sqrt{t^2+1}$. Then the relative speed of the fluid motion is
\[v_n=v-y'(t)=\frac{V(\lambda_0)+t\lambda_0}{\sqrt{t^2+1}}-\frac{\lambda_0t}{\sqrt{t^2+1}}=\frac{V(\lambda_0)}{\sqrt{t^2+1}}.\]
As $V>0$ the state before the jump is to the left of the state after the jump. The conditions on the hydraulic jump (mass and momentum conservation laws) are \cite{Stoker}:
\begin{equation}\label{RG}
[hv_n]=0,\;\;\;\left[hv_n+\frac{1}{2}h^2\right]=0,\;\;\;[u]=0.
\end{equation}
As usual, the square brackets denote the difference of the limiting values on the jump of the function in the states before and after the jump: $[f]=f_1-f_2$. Substitution of the representation \eqref{NHrepr} allows rewriting of conditions \eqref{RG} in the invariant form
\begin{equation}\label{RGI}
[HV]=0,\;\;\;[2HV^2+H^2]=0,\;\;\;[K]=0,\;\;\;[U]=0.
\end{equation}
Let $\sigma=HV=m/\cos\mu$. By virtue of the first condition of \eqref{RGI} we have $[\sigma]=0$. The second condition of \eqref{RGI} is convenient to rewrite as
\begin{equation}\label{RG2}
[\sigma/V^2+2V]=0.
\end{equation}
Function $\sigma/V^2+2V$ reaches its minimal value at $V=\sigma^{1/3}$, i.e. at the boundary of the domain of the solution (see \eqref{Vlim}). The flow is supercritical for $V>\sigma^{1/3}$, and subcritical for $V<\sigma^{1/3}$.

The jump stability condition is that in the state after the jump the fluid depth should be larger than the one in the state before the jump \cite{Stoker}. In terms of the solution under investigation this implies $H_2>H_1$ or, due to $\sigma$ conservation, $V_2<V_1$. This means that the state before the jump is supercritical, and the state after the jump is subcritical. This stability condition conforms with the Lax evolutionary condition for the strong discontinuity \cite{Lax}. Indeed, graphics of characteristics in figure \ref{f67} demonstrate, that incoming to the discontinuity $\lambda=\lambda_0$ are 3 characteristics from the left-hand side and 1 characteristic from the right-hand side. The outgoing ones are two characteristics from the right-hand side of the jump. Thus, the number of outgoing characteristics is one less then the number of conditions \eqref{RG} on the brake, which coincide with the Lax evolutionary condition.

According to the characteristics equations \eqref{charact1}, function $\mu-\tau$ is Lagrange invariant, i.e. it conserves along particle's trajectory. As the trajectory of each particle is continuous at the hydraulic jump, values of $\mu$ are the same on both sides on the jump. By virtue of $[\sigma]=0$, constant $m$ also conserves on the discontinuity: $[m]=0$. Thus, the third condition of \eqref{RGI} is satisfied. The last condition of \eqref{RGI} is equivalent to the demand of continuity on the jump of function $f$ in expression for the velocity component $U$.

Let the state ``1'' before the jump be known (i.e., constants $m$ and $b_1$ are fixed and the solution is constructed according to  formulae above), and the position of the jump $\lambda=\lambda_0$ is specified. It is required to determine the state ``2'' after the jump, i.e. to select constant $b_2\ne b_1$, such that the solution ``2'' conjugates with the solution ``1'' by conditions \eqref{RGI}.

The algorithm of solutions conjugation is the following. Note that the solution $\lambda(\mu)$ of equation \eqref{ImplEqn} is a monotonic function. Hence, value $\mu=\mu_1$ in the state ``1'' on the hydraulic jump can be uniquely found provided the jump position $\lambda=\lambda_0$ is given. Known constants $m$ and $\mu_1$ allow determination of $\sigma=m/\cos\mu_1$. Next, for known relative flow velocity $V_1$ one can determine the velocity $V_2$ after the jump from condition \eqref{RG2}. As $\sigma$ conserves on the brake, equation \eqref{ImplEqn} gives constant $b_2$ in the state after the discontinuity. Let us show that obtained values of $\sigma$, $V_2$, and $b_2$ satisfy condition \eqref{Ddef} of existence of physically meaningful solution of equation \eqref{ImplEqn}. Indeed,
\[{\cal D}=\sigma^2-\left(\frac{b^2-\lambda^2}{3}\right)^3=\sigma^2-\left(\frac{V^2}{3}+\frac{2\sigma}{3V}\right)^3.\]
Function $V^2+2\sigma/V$ reaches its minimal value $3\sigma^{2/3}$ at $V=\sigma^{1/3}$, which corresponds to the discriminant curve ${\cal D}=0$. For all another values of $V$ this function exceeds its minimal value, which guarantees the desired inequality ${\cal D}<0$ satisfaction. Finally, the solution behind the jump is constructed by finding the subcritical solution of equation \eqref{ImplEqn} with initial data $\lambda(\mu_1)=\lambda_0$.

\mysection{The general description of motion} Let us give an interpretation of the fluid motion governed by the obtained solution. For the fixed integral curve of equation \eqref{ImplEqn} the motion takes place in a stripe, bounded by linear source and drain of fluid in positions $y=\lambda_\ast\sqrt{t^2+1}$, and $y=\lambda^\ast\sqrt{t^2+1}$. Both lines are envelopes of the $C_-$ characteristic family of equations \eqref{shall2D}. Along $Ox$ axis the stripe is not limited. However, due to the linear dependence of the velocity component $u$ on the Cartesian coordinate $x$, the stripe should be limited with the use of, for example, two pistons. Thus, the flow takes place in a pool, whose two opposite walls are moving source and drain of fluid and another two walls are moving pistons. The free surface of fluid is convex upwards on subcritical flows, and is convex downwards on supercritical ones. The hydraulic jump along a moving line  $y=y_0\sqrt{t^2+1}$ is possible. In this case the convex downwards free surface of fluid switches stepwise on the jump to the convex upwards free surface.

The particle trajectories are determined as the solution of Cauchy problem
\[
\frac{dx}{dt}=u,\;\;\;\frac{dy}{dt}=v,\;\;\;x|_{t=0}=x_0,\;\;\;y|_{t=0}=y_0.
\]
Substitution of the obtained solution and integration gives the following
\[\mu(\lambda)-\arctan t=c_1,\;\;\;x=(c_2+f(c_1)\arctan t)\sqrt{t^2+1}/\cos\mu.\]
Here the dependence $\mu(\lambda)$ is defined by the chosen integral curve of equation \eqref{ImplEqn}. This dependence is monotonic, and, hence, is invertible at $\lambda_\ast\le\lambda\le\lambda^\ast$. Let the value $\lambda=\lambda_0=y_0$ corresponds to $\mu=\mu_0$. Then equations of the particle trajectory, starting from the position $(x_0,y_0)$ at $t=0$ are given by formulae
\begin{equation}\label{traject}
x=\bigl(x_0\cos\mu_0+f(\mu_0)\arctan t\bigr)\sqrt{t^2+1}/\cos\mu,\;\;\;
\mu(\lambda)=\mu_0+\arctan t,\;\;\;y=\lambda\sqrt{t^2+1}.
\end{equation}
Lines $y=\mathrm{const}$ are frozen into the flow, i.e. they consist of the same particles at all moments of time. The fluid flow along $Oy$ axis is fixed and is completely determined by the chosen integral curve of equation \eqref{ImplEqn}. Fluid motion along $Ox$ axes can be modified using the arbitrariness of function $f$.

The completeness of the analytical investigation of the solution gives opportunity to utilize it for the testing numerical solvers of hyperbolic systems of equations with several independent variables \cite{LeVeque}.

The work was supported by RFBR (project 05-01-00080), President programme of support of leading scientific schools and young scientists (grants Sc.Sch.-5245.2006.1, MK-1521.2007.1), and by Integration project 2.15 of Siberian Branch of RAS.


\begin{thebibliography}{99}
\bibitem{LVO} Ovsyannikov L V 1982  {\it Group analysis of differential equations} (New York: Academic Press).

\bibitem{LVO1} Ovsyannikov L V 1958 Groups and invariant-group solutions of differential equations. (in Russian) {\it Dokl. Akad. Nauk SSSR} {\bf 118} 439--442.

\bibitem{Pavl} Pavlenko A S 2005 Symmetries and solutions of equations of two-dimensional motions of politropic gas {\it Siberian Electronic Mathematical Reports} (http://semr.math.nsc.ru) {\bf 2} 291--307

\bibitem{Arnold} Arnold V I 1983 {\it Geometrical Methods In The Theory Of Ordinary Differential Equations, Grundlehren der mathematischen Wissenschaften. V. 250} (New-York: Springer-Verlag)

\bibitem{Stoker} Stoker J J 1957 {\it Water Waves: The Mathematical Theory with Applications} (New York: Interscience)

\bibitem{Lax} Lax P S 1957 Hyperbolic systems of conservetion laws II {\it Comm. Pure Appl. Math.} {\bf 10}(4) 537--566.

\bibitem{LeVeque} LeVeque R J 2002 {\it Finite Volume Methods For Hyperbolic Problems. Texts in Applied Mathematics} (Cambridge: Cambridge University Press)
\end{thebibliography}
\end{document}